# Nanobubble size controls gas hydrate nucleation in supercooled water

Ramkhelavan Kanaujiya[a], Atanu K. Metya[c], Rajnish Kumar[a] and Tarak K Patra[a,b]∗

[a]Department of Chemical Engineering, Indian Institute of Technology Madras, Chennai, Tamil Nadu – 600036
[b]Center for Atomistic Modeling and Materials Design, Indian Institute of Technology Madras, Chennai, TN-600036, India
[c]Department of Chemical and Biochemical Engineering, Indian Institute of Technology Patna, Patna – 801106, India

**ABSTRACT**

Gas hydrates are crystalline compounds formed when water ($H_2O$) molecules encapsulate guest gas molecules under high pressure and low temperatures. They have gained significant interest due to their potential as alternative energy resources and their applications in gas storage, transportation, and carbon sequestration. However, the fundamental mechanisms governing their formation, especially the influence of gas bubbles, remain poorly understood. In this study, we use molecular dynamics (MD) simulations to examine how methane ($CH_4$) nanobubble size modulates hydrate formation in supercooled water. Nanobubbles of different sizes are generated by modulating the $CH_4$ concentration in a $CH_4$–$H_2O$ mixture during equilibration under high-temperature and low-pressure conditions, followed by quenching to low temperature and high pressure to induce gas hydrate nucleation and subsequent growth. The simulations reveal a strong correlation between nanobubble size and the extent of hydrate formation. Specifically, the extent of hydrate formation increases with bubble size in the small-to-intermediate regime. However, beyond a critical bubble size threshold, the hydrate formation efficiency declines. The work provides new molecular-level insight into how nanobubble size modulates gas hydrate nucleation and growth dynamics.

## 1. INTRODUCTION

Gas hydrates, also known as clathrate hydrates, are crystalline compounds in which hydrogen-bonded water molecules form cage-like structures that encapsulate "guest" gas molecules such as carbon dioxide ($CO_2$), methane ($CH_4$), nitrogen ($N_2$), hydrogen ($H_2$), and other light hydrocarbons under conditions of high pressure and low temperature [1–3]. These hydrates are predominantly found in deep-sea sediments and permafrost regions and have drawn significant interest as potential alternative energy resources [4]. Beyond energy, gas hydrates have applications in gas separation and storage [5–8]. Conversely, hydrates can form in oil and gas pipelines and cause transport blockages, creating major flow-assurance and safety concerns. Therefore, developing strategies to control gas hydrate formation remains a central ongoing research focus—both to promote hydrate formation in applications such as gas storage, carbon capture, and sequestration, and to inhibit unwanted hydrate formation in subsea pipelines and processing facilities. Advances in the molecular-level understanding and engineering of the gas-water interface continue to drive this field forward [9–13].

Gas hydrate crystals are commonly classified as Structure I (sI), Structure II (sII), and Structure H (sH), which are assembled by the arrangement of polyhedral water cages $5^{12}$, $5^{12}6^{2}$, and $5^{12}6^{4}$, respectively. Each cage is capable of stabilizing guest molecules according to their size, shape, and chemical nature [14,15]. The formation of gas hydrates generally proceeds through two stages: nucleation and growth [16,17]. The nucleation stage is a rare stochastic event that poses greater experimental challenges than the subsequent growth stage, as it involves the spontaneous emergence of a new phase at the nanoscale [18–23]. Several hypotheses have been proposed to elucidate the microscopic mechanisms governing the nucleation of clathrate hydrates. The first hypothesis suggests that cage-like water networks, resembling the polyhedral cavities of clathrate structures, are pre-organized around guest molecules in solution [24]. These transient, labile clusters subsequently associate within the liquid phase to generate stable nuclei [25,26]. Once the cluster size exceeds a critical threshold, spontaneous hydrate crystal growth occurs. The formation and evolution of these early clusters are believed to play a fundamental role in determining nucleation and growth kinetics [27,28]. An alternative hypothesis emphasizes the emergence of local structural order among guest molecules. In this view, gas molecules first adopt a locally ordered arrangement; when their number exceeds that required for a critical nucleus, surrounding water molecules reorganize to establish the clathrate framework [29]. Another proposed mechanism involves a two-step "blob" pathway, in which gas molecules initially form a locally dense region that evolves into an amorphous clathrate nucleus, which then transforms into a crystalline structure [30,31]. Simulation studies conducted across various time and length scales indicate that hydrate nucleation may proceed through multiple pathways, involving either metastable intermediates or the direct emergence of a thermodynamically stable phase [32–34]. Collectively, these findings suggest that hydrate crystallization is a multistep process in which amorphous or intermediate structures evolve into crystalline cages, while fluctuations in local gas concentration and water ordering critically influence the likelihood of nucleation [15,35,36]. Recent molecular dynamics (MD) studies further indicate that pre-nucleation gas nanobubbles–often spherical or cylindrical–form prior to hydrate nucleation. The curvature of the gas–liquid interface thus plays a vital role in nucleation, depending strongly on the gas concentration in the aqueous phase [37–41]. Under ambient conditions, $CH_4$ is only sparingly soluble in water. When introduced, it readily forms nanobubbles—gas-filled cavities dispersed in water with diameters ranging from a few nanometers to ~1000 nm [42]. Methane concentration strongly influences nanobubble stability; lower $CH_4$ levels lead to smaller, more mobile bubbles with higher diffusion coefficients, whereas bulk nanobubbles require oversaturation to remain stable.[42] Stable nanobubbles are observed to emerge during hydrate dissociation[43] and can reform hydrates upon re-exposure to suitable conditions. [44]

A central question in controlling hydrate formation is therefore: How do $CH_4$ nanobubbles transform into stable hydrate nuclei under high-pressure/low-temperature conditions? Elucidating this pathway is critical because it governs nucleation kinetics, interfacial stability, and subsequent hydrate growth. Although the thermodynamics and kinetics of hydrate formation have been widely explored, the influence of nanobubble size during nucleation remains poorly quantified. Bubble size alters interfacial curvature and Laplace pressure, affecting gas dissolution, local supersaturation, and the stability of nascent clathrate clusters. Consequently, nanobubble size can exert a decisive influence on the nucleation pathway. However, the quantitative correlation between bubble size and hydrate nucleation remains elusive, limiting our ability to predict and control hydrate formation mechanisms.

In this study, we address the above gap by performing MD simulations to explore how $CH_4$ bubble size influences gas hydrate formation. We vary the $CH_4$ content in $CH_4$–$H_2O$ mixtures to achieve a distribution of bubble sizes and examine its effects on the gas hydrate formation. We estimate the four-body order parameter ($F_4$) of water over a 1000 ns trajectory under low-temperature and high-pressure conditions that are conducive to hydrate formation. The simulations suggest a strong correlation between the bubble size distribution and the $F_4$ order parameter, likely due to an enhanced nucleation probability resulting from an optimal local gas density [45] [46]. However, beyond a certain threshold size, the hydrate formation rate begins to decrease. By establishing how gas nanobubble size correlates with hydrate nucleation under controlled conditions, this work provides molecular-level insights that are useful for the design of more efficient and tunable hydrate systems.

## 2. MODEL AND METHODOLOGY

We employ the OPLS-AA force field for $CH_4$ [47] and the TIP4P/ice model for water [48]. A schematic representation of the gas and water models is shown in Figure 1a. In the TIP4P/ice water model, each water molecule is treated as a rigid body with fixed geometry enforced through the LINCS constraint algorithm [49], whereas a $CH_4$ molecule is represented as a flexible molecule described by the OPLS-AA force field. The cross-interaction parameters for dissimilar species are derived using the geometric mean combination rule [50]. Non-bonded short-range interactions, i.e., the van der Waals and Coulombic interactions, are truncated at a distance of 1.2 nm. The long-range electrostatic interaction is computed via the particle mesh Ewald (PME) method, employing a grid spacing of 0.1 nm and a real-space cutoff of 1.2 nm [51]. The initial systems are prepared by randomly inserting molecules into a cubic box of dimensions 5 × 5 × 5 nm³, as shown in Figure 1b. Periodic boundary conditions are applied in all spatial dimensions. The total number of water molecules is 2,944, and the $CH_4$ content varies from 5 to 20 mol%. The equations of motion are integrated using the leapfrog algorithm [52] with a time step of 2 fs. Initial energy minimization is performed using the steepest descent algorithm [53]. Subsequently, we equilibrate the system in an isothermal–isobaric ensemble (NPT) at 300 K and 10 MPa for 10 ns. Production runs are carried out in the NPT ensemble at 240 K and 50 MPa for 1 μs. The temperature and pressure are maintained using the Nosé-Hoover thermostat [54,55] and the Parrinello–Rahman barostat [56], respectively. All simulations are conducted using the GROMACS package (version 2023) [57]. We perform three independent simulations for each system, starting with different initial configurations, and report data averaged over these replicas.

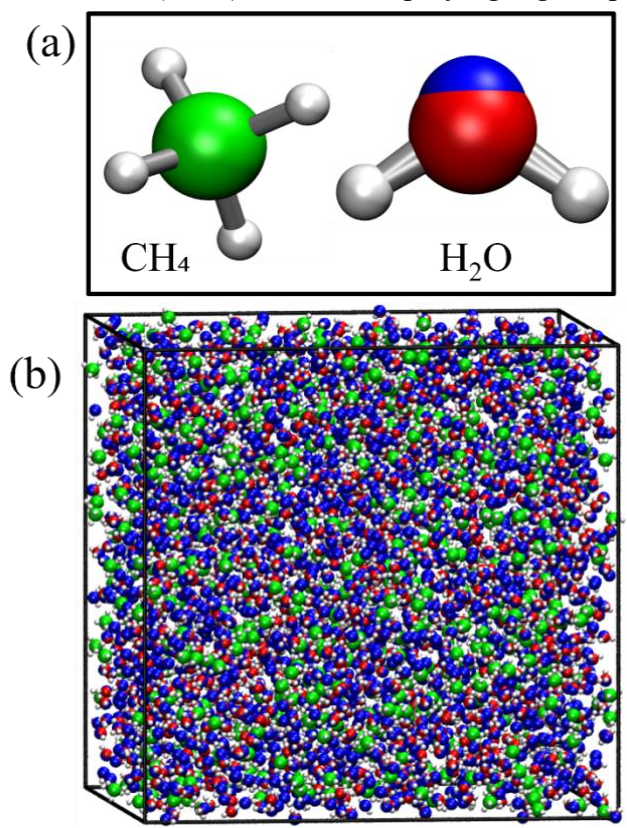

*Figure 1: Panel (a) shows a schematic representation of $CH_4$ and $H_2O$ models, while Panel (b) depicts the initial configuration of the system containing 20 mol % $CH_4$. Carbon atoms of methane are shown in green, oxygen atoms of water in blue, and the virtual site of water in red. All hydrogen atoms associated with both molecules are represented by white spheres.*

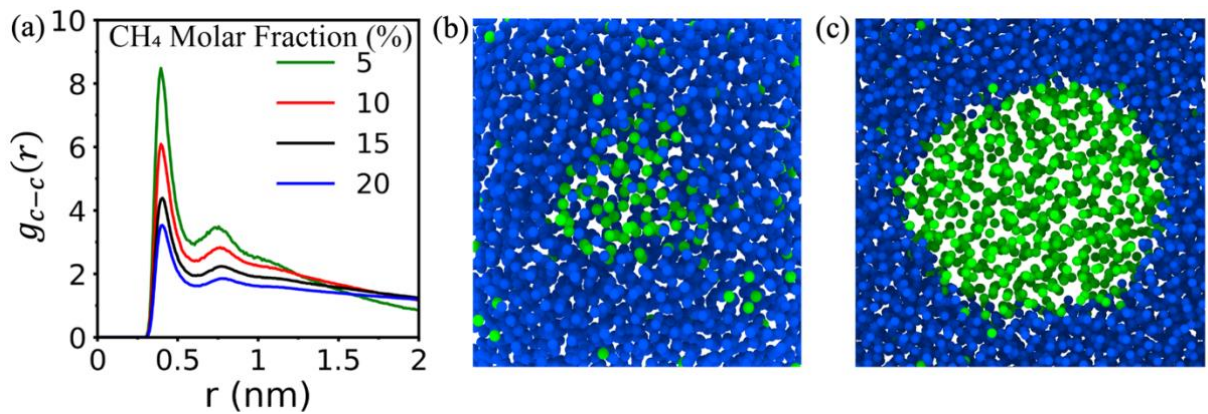

*Figure 2: The carbon-carbon RDFs for different gas concentrations in water are shown in (a). The RDF is calculated based on the last 2 ns trajectory of the equilibration run. The MD snapshot of the system after the equilibration run are shown in (b) and (c) for gas concentration 5% and 20%, respectively. The gas and water are represented by green and blue colors in (b) and (c), respectively.*

## 3. RESULTS AND DISCUSSION

Gas bubbles are formed during the equilibration at 300 K and 10 MPa. We compute the methane carbon–carbon radial distribution function (RDF). Figure 2 presents the RDF profiles for $CH_4$ contents of 5, 10, 15, and 20 mol%. A pronounced first peak appears around 0.45 nm for all the cases, which corresponds to the methane–methane separation distance within the first coordination shell. The first peak height is largest at low $CH_4$ concentration and reduces with increasing concentration, consistent with relatively smaller bubbles at low $CH_4$ and significantly larger bubbles at high $CH_4$ content in the $CH_4$–$H_2O$ mixture. This trend agrees with the MD snapshots in Figure 2(b-c). Equilibrated systems are subjected to a production run at 240 K and 50 MPa, during which hydrates nucleate and grow. We quantify the number of nanobubbles as a function of size to understand their distribution and temporal evolution. Figures 3(a–f) present MD snapshots at 0 ns, 20 ns, and 1000 ns for systems containing 15 and 20 mol% $CH_4$. For 15 mol% $CH_4$, (Figures 3(a–c)), a distinct nanobubble formed during the equilibration (i.e., at high temperature and low pressure) gradually diminishes and eventually completely dissociates by 1000 ns. In contrast, for 20 mol% $CH_4$ (Figures 3(d–f)), bubble dissociation is markedly slow, and a finite-size nanobubble persists even at 1000 ns, indicating enhanced nanobubble stability at higher $CH_4$ concentrations in the $CH_4$–$H_2O$ mixture.

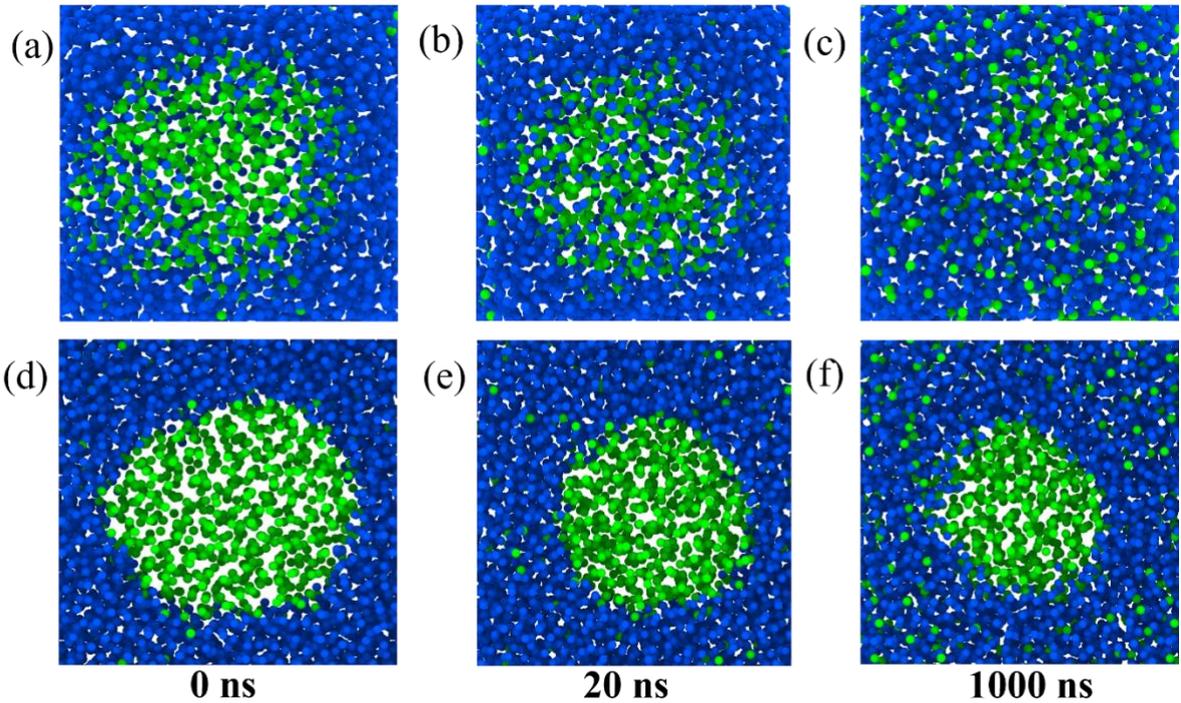

*Figure 3. Nanobubble dissociation. MD snapshots of the system illustrating the evolution of a nanobubble at 0 ns, 20 ns, and 1000 ns of the production run in panels (a, d), (b, e), and (c, f), respectively. Panels (a–c) correspond to 15 mol% $CH_4$, while panels (d–f) correspond to 20 mol% $CH_4$. The gas and water are represented by green and blue colors, respectively.*

To quantitatively assess the nanobubble evolution, Figures 4(a–d) present the temporal evolution of the nanobubble size distribution, expressed as the probability of observing a bubble containing $N_b$ gas molecules at different simulation times - 0 ns (green), 20 ns (red), and 1000 ns (blue). It suggests the distribution shifts toward smaller bubble sizes over time. In Figure 4(a) for 5 mol% $CH_4$ concentration, at the onset of the production simulation (0 ns), a nanobubble contains roughly 2 to 100 $CH_4$ molecules. It suggests the distribution shifts toward smaller bubble sizes over time. The largest bubble size decreases from roughly 100 $CH_4$ molecules to nearly 10 molecules by 1000 ns for a 5 mol% $CH_4$ concentration. At a 20 mol% $CH_4$ concentration, a single large bubble comprising ~1000 molecules is initially present, which decreases in size to ~950 molecules by the end of the production run. This dissociation leads to the formation of new bubbles of size less than 10 molecules. These results suggest bubbles of sizes 2 to 10 are preferred for most of the cases, whereas at very high $CH_4$ concentrations, large-size bubbles tend to persist in the system.

The hydrate formation is assessed through the structural ordering of water molecules using the four-body order parameter ($F_4$), which is defined as $F_4 = \left(\frac{1}{k}\right) \sum \cos 3\phi_i$. Here, $\phi_i$ is the H-O—O-H torsional angle and k is the number of oxygen-oxygen pairs within 0.35 nm of the selected molecule. The $F_4$ value is known to be around -0.4 for ice, -0.04 for liquid water, and approximately 0.7 for the hydrate structure [58]. These distinct values allow for the differentiation between the aqueous and hydrate phases. This serves as an effective indicator of clathrate-like ordering. We calculate the $F_4$ order parameter using the GRADE code [59]. The $F_4$ order parameter is calculated throughout the production run to quantify the nucleation and growth of methane hydrates at different $CH_4$ concentrations.

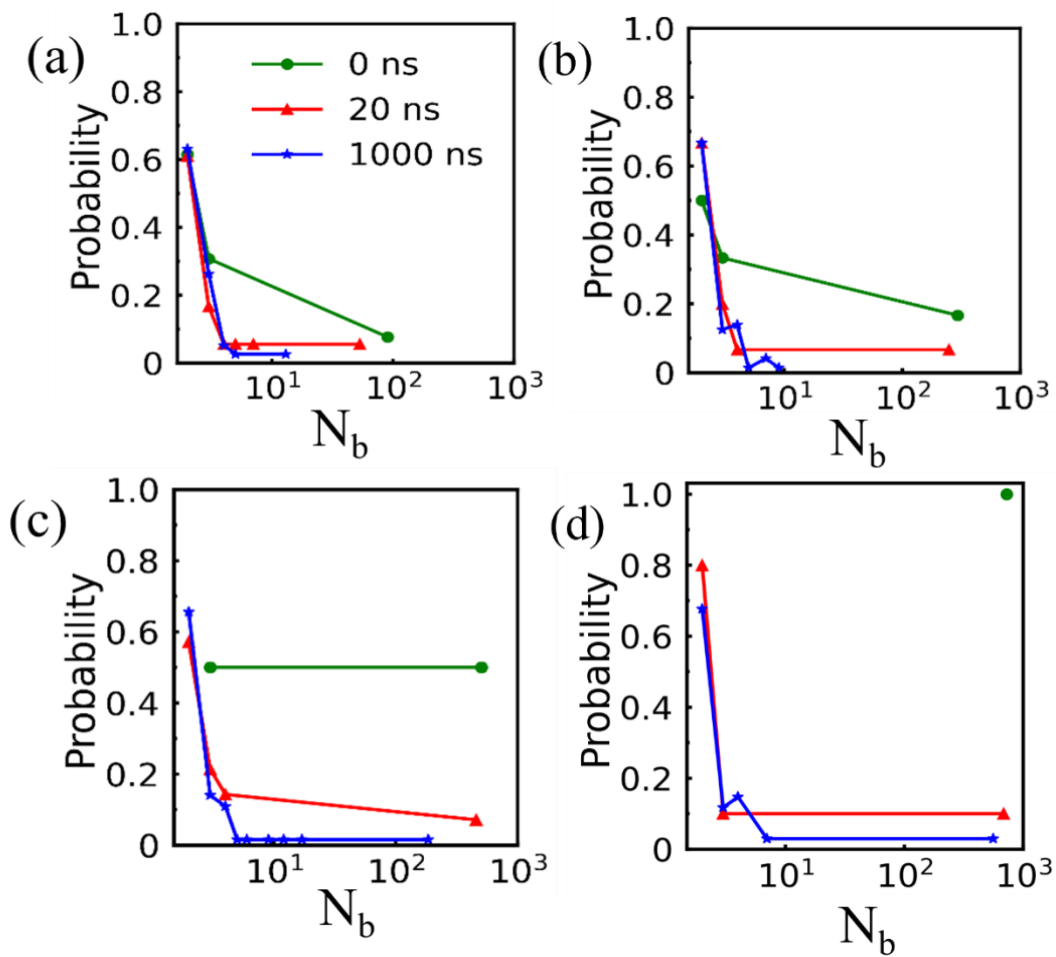

*Figure 4: Probability distributions of nanobubble size at three different simulation times (0, 20, and 1000 ns) of the production run. Panels (a), (b), (c), and (d) correspond to CH₄ molar fractions of 5%, 10%, 15%, and 20%, respectively.*

Figure 5(a) shows the $F_4$ as a function of time for the system containing 5, 10, 15, and 20 mol% of $CH_4$. At low (5 mol%) and high (20 mol%) $CH_4$, hydrate nucleation is very slow. Intermediate $CH_4$ contents (10 and 15 mol%) show clear nucleation after ~200 ns, followed by a steady rise in $F_4$, which indicates the gradual formation of hydrate-like structures. Figure 5(b) shows the $F_4$ of the system as a function of the average nanobubble size (both averaged over the last 2 ns). It suggests a relationship between the extent of hydrate formation and nanobubble size. Hydrate formation increases with bubble size, reaches a plateau, and then decreases with further increases in bubble size. This indicates that hydrate formation is most favorable at an intermediate nanobubble size, where the contact between methane and water molecules is most effective for nucleation. Even a smaller bubble size may have higher methane-water contact; however, below a certain threshold, perhaps sufficient methane molecules are not present for hydrate nucleation and growth.

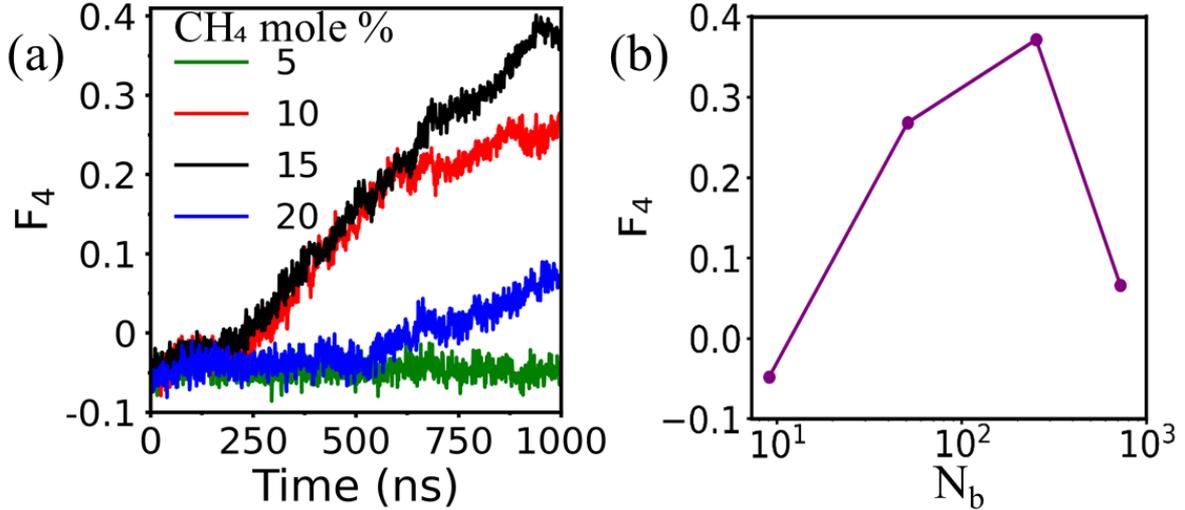

*Figure 2: The $F_4$ order parameter of the system as a function of simulation time is shown in (a). The $F_4$ order parameter is plotted as a function of nanobubble size in (b); both quantities are average over the final 2 ns of the production run.*

We further analyze different types of hydrate cages formed during the simulation by examining the connectivity of five- and six-membered rings based on the atomic coordinates of oxygen atoms from neighbouring water molecules. Using the GRADE code developed by Mahmoudinobar and Dias, we identify the three most common cages: small ($5^{12}$), large sI-type ($5^{12}6^2$), and some large sII-type ($5^{12}6^4$) [58]. This analysis provides a detailed characterization of the local water network and the structural arrangements that underpin methane hydrate cage formation. Figure 6(a–d) presents the number of each cage type over the course of the simulation. At 5 mol% $CH_4$ (Figure 6a), cage formation is slow and limited, predominantly producing small ($5^{12}$) cages. At 10 and 15 mol% $CH_4$ (Figures 6b and 6c), we observe faster and more consistent cage formation, with a large number of $5^{12}$ and $5^{12}6^2$ cages. At 20 mol%, Figure 6d indicates that a very high $CH_4$ content in $CH_4$–$H_2O$ mixture leads to defects and fewer cages. We observe that primarily $5^{12}$ cages formed during hydrate formation, and a relatively smaller number of $5^{12}6^2$ and $5^{12}6^4$ cages are observed in all our simulations. This is consistent with previously reported studies on hydrate nucleation [41]. It is important to note that a thermodynamically stable SI hydrate of methane should have 3 times as many $5^{12}6^2$ cages compared to $5^{12}$ cages in the unit cell crystal. Figure 7(a–d) shows MD snapshots of the systems at the end of production runs. These snapshots indicate that the cage formation generally increases with nanobubble size; however, when nanobubbles become excessively large, the total number of cages decreases. Overall, intermediate $CH_4$ concentrations appear to promote a large number of cage formations due to the presence of optimally sized nanobubbles.

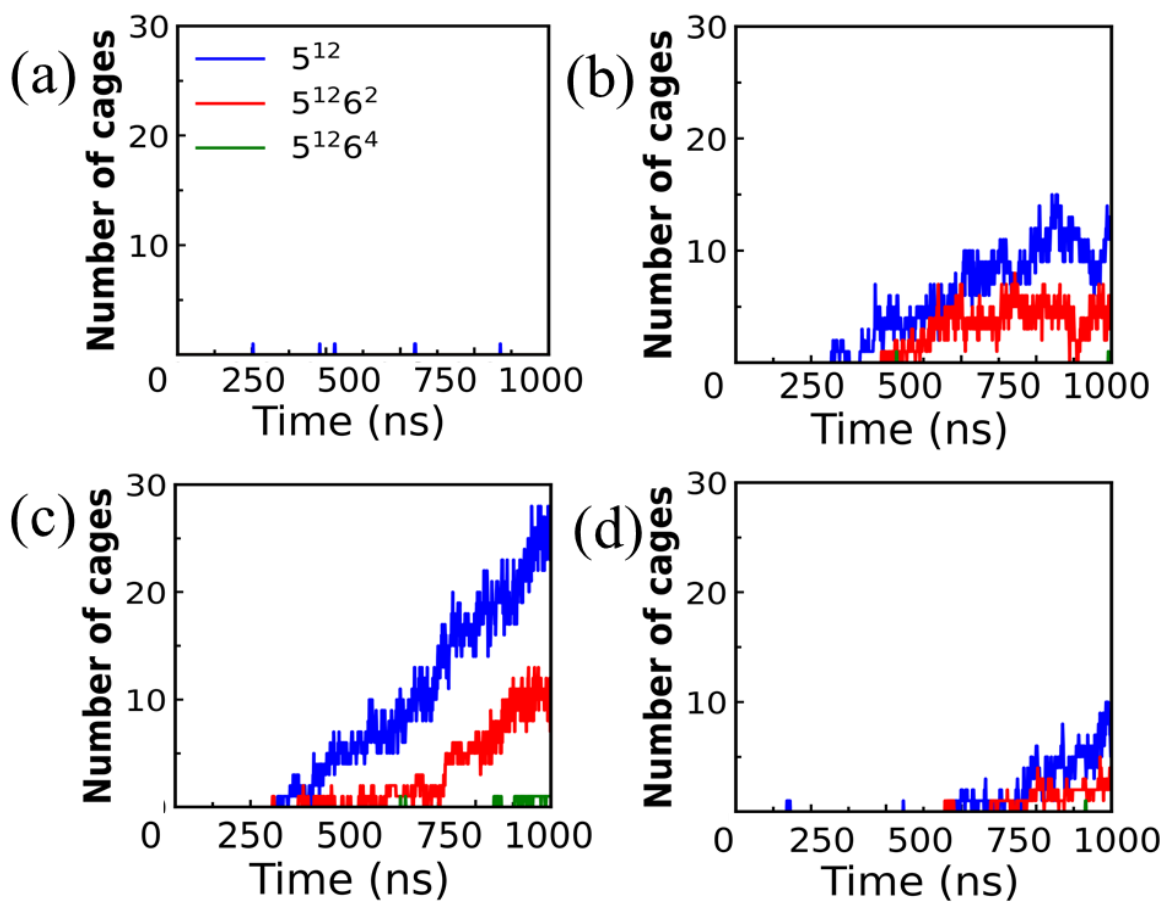

Figure 3: Time evolution of different types of cages: $5^{12}$ (blue), $5^{12}6^2$ (red), and $5^{12}6^4$ (green) during nucleation and growth of gas hydrate formation. Panels (a–d) correspond to $CH_4$ = 5, 10, 15, and 20 mol%, respectively.

## 4. CONCLUSIONS

Gas hydrates constitute an important class of crystalline compounds with significant potential in energy storage, gas transportation, and gas separation applications. Understanding and controlling the mechanisms governing gas hydrate formation are crucial for optimizing their utilization across diverse technological domains. In this study, MD simulations are performed to investigate how $CH_4$ content and the resulting nanobubble size affect methane hydrate nucleation and growth. The gas–water mixture, prepared with different gas concentrations, is initially equilibrated under high-temperature and low-pressure conditions. Upon quenching to low temperature and high pressure, the system undergoes structural transformations that are monitored over a microsecond timescale to capture the evolution of hydrate formation. Our results show that increasing $CH_4$ concentration in the mixture shifts the equilibrium towards larger size nanobubble formation. The probability of hydrate nucleation increases with bubble size, reaching a maximum at a critical threshold beyond which hydrate nucleation becomes unfavourable. The critical threshold corresponds to ~15 mol% of $CH_4$ in the solution. A larger $CH_4$ nanobubble appears to disrupt the hydrogen-bonded network that is required for stable clathrate cage formation. Analysis of the $F_4$ order parameter, cage counting, and radial distribution functions supports this observation. Moreover, nanobubble analysis and MD snapshots demonstrate that the dynamics of $CH_4$ nanobubble dissociation strongly influence the nucleation process. These findings highlight a delicate balance between gas availability and structural disruption in hydrate systems: an optimal number of guest molecules in nanobubbles that promotes nucleation, while excessive $CH_4$ leads to supersaturation, large-sized nanobubbles, and reduced ordering. Overall, our study provides key insights into the microscopic mechanisms governing hydrate formation, suggesting that controlling nanobubble size and distribution could be an effective strategy for tuning hydrate nucleation kinetics in applications such as gas storage, flow assurance, and carbon sequestration.

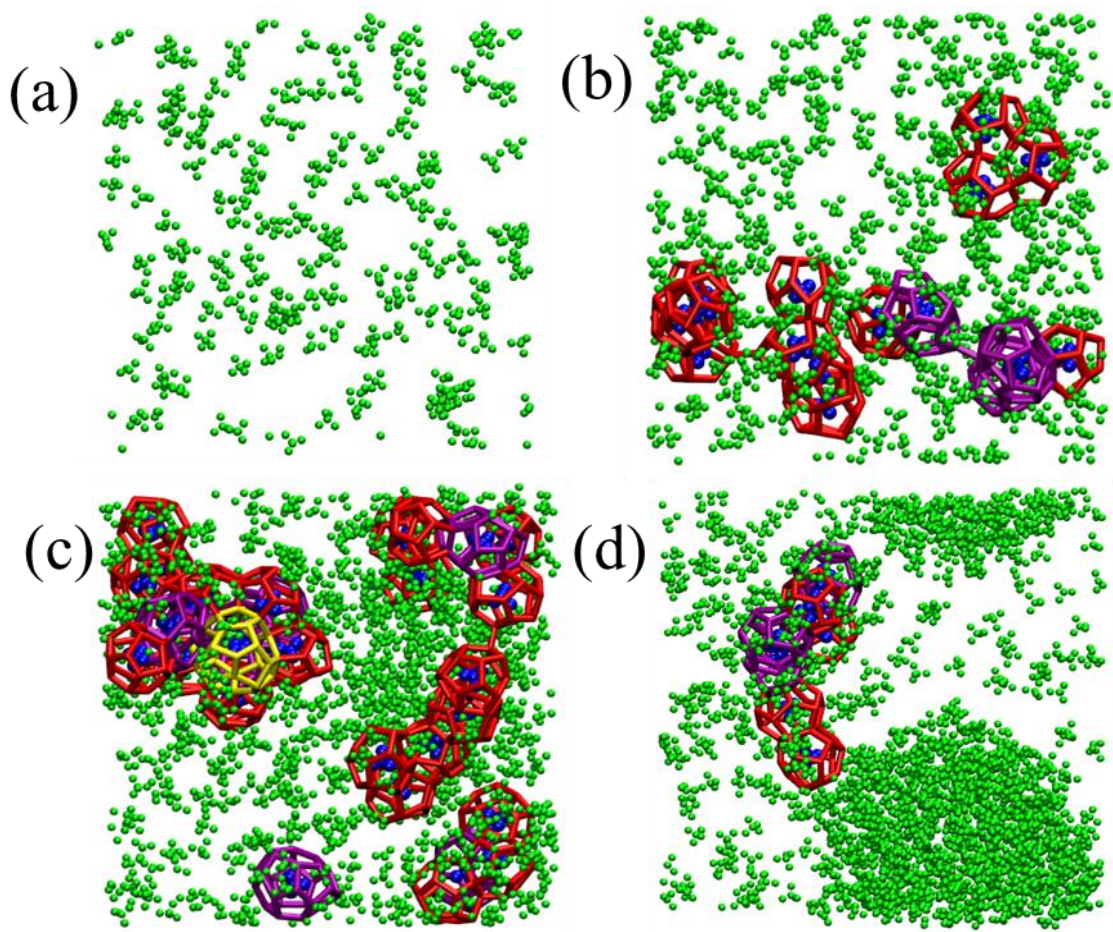

*Figure 4: MD snapshots at the end of production runs. Water and methane are shown in blue and green particles, respectively. Water molecules that do not participate in cage formation are omitted for better visual clarity. Cages $5^{12}$, $5^{12}6^2$, and $5^{12}6^4$ are depicted in red, purple, and yellow, respectively. Panels (a-d) correspond to 5, 10, 15, and 20 mol% $CH_4$, respectively.*

**ACKNOWLEDGEMENTS**

We acknowledge the use of the computing resources at HPCE, IIT Madras. TKP acknowledges SERB(ANRF) for support through a core research grant (CRG/2022/006926).

**REFERENCES**

1. S. S. Byk and V. I. Fomina, Gas Hydrates, *Russ. Chem. Rev.*, 1968, **37**, 469.
2. E. D. S. Jr, C. A. Koh and C. A. Koh, *Clathrate Hydrates of Natural Gases*, CRC Press, Boca Raton, 3rd edn., 2007.
3. P. Linga, R. Kumar and P. Englezos, Gas hydrate formation from hydrogen/carbon dioxide and nitrogen/carbon dioxide gas mixtures, *Chem. Eng. Sci.*, 2007, **62**, 4268–4276.
4. E. D. Sloan Jr., Natural Gas Hydrates, *J. Pet. Technol.*, 1991, **43**, 1414–1417.
5. C. A. Koh, Towards a fundamental understanding of natural gas hydrates, *Chem. Soc. Rev.*, 2002, **31**, 157–167.
6. J. Rajnauth, M. Barrufet and G. Falcone, Potential Industry Applications Using Gas Hydrate Technology, *West Indian J. Eng.*, DOI:10.2118/133466-MS.
7. A. Eslamimanesh, A. H. Mohammadi, D. Richon, P. Naidoo and D. Ramjugernath, Application of gas hydrate formation in separation processes: A review of experimental studies, *J. Chem. Thermodyn.*, 2012, **46**, 62–71.
8. C. A. Koh, E. D. Sloan, A. K. Sum and D. T. Wu, Fundamentals and Applications of Gas Hydrates, *Annu. Rev. Chem. Biomol. Eng.*, 2011, **2**, 237–257.
9. J. A. Ripmeester and S. Alavi, Some current challenges in clathrate hydrate science: Nucleation, decomposition and the memory effect, *Curr. Opin. Solid State Mater. Sci.*, 2016, **20**, 344–351.
10. A. Farhadian, A. Shadloo, X. Zhao, R. S. Pavelyev, K. Peyvandi, Z. Qiu and M. A. Varfolomeev, Challenges and advantages of using environmentally friendly kinetic gas hydrate inhibitors for flow assurance application: A comprehensive review, *Fuel*, 2023, **336**, 127055.
11. Z. Chen, A. Farhadian, P. Naeiji, D. A. Martyushev and C. Chen, Molecular-Level insights into kinetic and agglomeration inhibition mechanisms of structure I and II gas hydrate formation, *Chem. Eng. J.*, 2025, **511**, 162194.
12. U. Tsutomu, S. Ren, Y. Kenji and G. Kazutoshi, Nucleation Behavior of Single-Gas Hydrates in the Batch-type Stirred Reactor and Their Promotion Effect with Ultrafine Bubbles: A Mini Review and Perspectives, *Energy Fuels*, 2022, **36**, 10444–10457.
13. T. Uchida, M. Hayama, M. Oshima and K. Yamazaki, Nucleation Probability of Methane + Propane Mixed-Gas Hydrate Depending on Gas Composition, *Energy Fuels*, 2025, **39**, 4782–4789.
14. K. A. Kvenvolden, A review of the geochemistry of methane in natural gas hydrate, *Org. Geochem.*, 1995, **23**, 997–1008.
15. Z. He, K. M. Gupta, P. Linga and J. Jiang, Molecular insights into the crystal nucleation and growth of CH4 and CO2 mixed hydrates from microsecond simulations, *J. Phys. Chem. C*, 2016, **120**, 25225–25326.
16. B. Kvamme, S. A. Aromada, N. Saeidi, T. Hustache-Marmou and P. Gjerstad, Hydrate Nucleation, Growth, and Induction, *ACS Omega*, 2020, **5**, 2603–2619.
17. D. Kashchiev and A. Firoozabadi, Nucleation of gas hydrates, *J. Cryst. Growth*, 2002, **243**, 476–489.
18. W. Ke, T. M. Svartaas and D. Chen, A review of gas hydrate nucleation theories and growth models, *J. Nat. Gas Sci. Eng.*, 2019, **61**, 169–196.
19. L. C. Jacobson and V. Molinero, Can Amorphous Nuclei Grow Crystalline Clathrates? The Size and Crystallinity of Critical Clathrate Nuclei, *J. Am. Chem. Soc.*, 2011, **133**, 6458–6463.

20 J. Lee, J. Yang, S. G. Kwon and T. Hyeon, Nonclassical nucleation and growth of inorganic nanoparticles, *Nat. Rev. Mater.*, 2016, **1**, 1–16.
21 J. Li and F. L. Deepak, In Situ Kinetic Observations on Crystal Nucleation and Growth, *Chem. Rev.*, 2022, **122**, 16911–16982.
22 Y. Lu, Y. Feng, D. Guan, X. lv, Q. Li, L. Zhang, J. Zhao, L. Yang and Y. Song, A molecular dynamics study on nanobubble formation and dynamics via methane hydrate dissociation, *Fuel*, 2023, **341**, 127650.
23 L.-D. Shiau, Modelling of the Polymorph Nucleation Based on Classical Nucleation Theory, *Crystals*, 2019, **9**, 69.
24 G.-J. Guo and Z. Zhang, Open questions on methane hydrate nucleation, *Commun. Chem.*, 2021, **4**, 102.
25 R. Christiansen and E. SLOAN, Mechanisms and Kinetics of Hydrate Formation, *Ann. N. Y. Acad. Sci.*, 2006, **715**, 283–305.
26 E. Sloan and F. Fleyfel, A Molecular Mechanism for Gas Hydrate Nucleation from Ice, *AIChE J.*, 1991, **37**, 1281–1292.
27 T. Uchida, K. Yamazaki and K. Gohara, Gas Nanobubbles as Nucleation Acceleration in the Gas-Hydrate Memory Effect, *J. Phys. Chem. C*, 2016, **120**, 26620–26629.
28 Y. Feng, Y. Han, P. Gao, Y. Kuang, L. Yang, J. Zhao and Y. Song, Study of hydrate nucleation and growth aided by micro-nanobubbles: Probing the hydrate memory effect, *Energy*, 2024, **290**, 130228.
29 R. Radhakrishnan and B. L. Trout, A new approach for studying nucleation phenomena using molecular simulations: Application to $CO_2$ hydrate clathrates, *J. Chem. Phys.*, 2002, **117**, 1786–1796.
30 C. Moon, P. C. Taylor and P. M. Rodger, Clathrate nucleation and inhibition from a molecular perspective, *Can. J. Phys.*, 2003, **81**, 451–457.
31 C. Moon, P. C. Taylor and P. M. Rodger, Molecular Dynamics Study of Gas Hydrate Formation, *J. Am. Chem. Soc.*, 2003, **125**, 4706–4707.
32 L. C. Jacobson, W. Hujo and V. Molinero, Amorphous Precursors in the Nucleation of Clathrate Hydrates, *J. Am. Chem. Soc.*, 2010, **132**, 11806–11811.
33 M. R. Walsh, G. T. Beckham, C. A. Koh, E. D. Sloan, D. T. Wu and A. K. Sum, Methane Hydrate Nucleation Rates from Molecular Dynamics Simulations: Effects of Aqueous Methane Concentration, Interfacial Curvature, and System Size, *J. Phys. Chem. C*, 2011, **115**, 21241–21248.
34 Z. Zhang, M. Walsh and G.-J. Guo, Microcanonical molecular simulations of methane hydrate nucleation and growth: Evidence that direct nucleation to sI hydrate is among the multiple nucleation pathways, *Phys Chem Chem Phys*, DOI:10.1039/C5CP00098J.
35 Y. Lu, X. Lv, Q. Li, L. Yang, L. Zhang, J. Zhao and Y. Song, Molecular behavior of hybrid gas hydrate nucleation: separation of soluble $H_2S$ from mixed gas, *Phys. Chem. Chem. Phys.*, 2022, **24**, 9509–9520.
36 L. C. Jacobson, W. Hujo and V. Molinero, Nucleation Pathways of Clathrate Hydrates: Effect of Guest Size and Solubility, *J. Phys. Chem. B*, 2010, **114**, 13796–13807.
37 M. R. Walsh, C. A. Koh, E. D. Sloan, A. K. Sum and D. T. Wu, Microsecond simulations of spontaneous methane hydrate nucleation and growth, *Science*, 2009, **326**, 1095–1098.
38 B. Li, W. Xiang, X. Dou, Y. Wu, W. Zhang, Z. Wang and J. Wang, Coarse-Grained Molecular Dynamics Simulation of Nucleation and Stability of Electrochemically Generated Nanobubbles, *Langmuir*, 2025, **41**, 8497–8509.
39 Z. He, P. Linga and J. Jiang, What are the key factors governing the nucleation of $CO_2$ hydrate?, *Phys. Chem. Chem. Phys.*, 2017, **19**, 15657–15661.
40 P. Kanani, K. Reddy, M. Adil and A. Metya, Impact of antifreeze and promoter on clathrate hydrate nucleation and growth, *J. Mol. Liq.*, 2025, **424**, 127118.

41 R. Kanaujiya, A. Metya, N. Choudhary, R. Kumar and T. Patra, Molecular Dynamics Insights into Tetrahydrofuran-Assisted Formation of $CH_4$, $CO_2$, and $H_2$ Gas Hydrates, *Phys. Chem. Chem. Phys.*, DOI:10.1039/D5CP01574J.

42 Y. Lu, L. Yang, Y. Kuang, Y. Song, J. Zhao and A. K. Sum, Molecular simulations on the stability and dynamics of bulk nanobubbles in aqueous environments, *Phys. Chem. Chem. Phys.*, 2021, **23**, 27533–27542.

43 S. A. Bagherzadeh, S. Alavi, J. Ripmeester and P. Englezos, Formation of methane nano-bubbles during hydrate decomposition and their effect on hydrate growth, *J. Chem. Phys.*, 2015, **142**, 214701.

44 T. Uchida, K. Yamazaki and K. Gohara, Gas Nano-Bubbles as Nucleation Acceleration in the Gas-Hydrate Memory Effect, *J. Phys. Chem. C*, DOI:10.1021/acs.jpcc.6b07995.

45 M. R. Walsh, G. T. Beckham, C. A. Koh, E. D. Sloan, D. T. Wu and A. K. Sum, Methane Hydrate Nucleation Rates from Molecular Dynamics Simulations: Effects of Aqueous Methane Concentration, Interfacial Curvature, and System Size, *J. Phys. Chem. C*, 2011, **115**, 21241–21248.

46 K.-F. Yan, X.-S. Li, Z.-Y. Chen, Z.-M. Xia, C.-G. Xu and Z. Zhang, Molecular Dynamics Simulation of the Crystal Nucleation and Growth Behavior of Methane Hydrate in the Presence of the Surface and Nanopores of Porous Sediment, *Langmuir*, 2016, **32**, 7975–7984.

47 W. L. Jorgensen, D. S. Maxwell and J. Tirado-Rives, Development and Testing of the OPLS All-Atom Force Field on Conformational Energetics and Properties of Organic Liquids, *J. Am. Chem. Soc.*, 1996, **118**, 11225–11236.

48 J. L. F. Abascal, E. Sanz, R. Fernández and C. Vega, A potential model for the study of ices and amorphous water: TIP4P/Ice, *J. Chem. Phys.*, 2005, **122**, 234511.

49 B. Hess, H. Bekker, H. J. C. Berendsen and J. G. E. M. Fraaije, LINCS: A linear constraint solver for molecular simulations, *J. Comput. Chem.*, 1997, **18**, 1463–1472.

50 M. P. Oliveira and P. H. Hünenberger, Influence of the Lennard-Jones Combination Rules on the Simulated Properties of Organic Liquids at Optimal Force-Field Parametrization, *J. Chem. Theory Comput.*, 2023, **19**, 2048–2063.

51 T. Darden, D. York and L. Pedersen, Particle mesh Ewald: An N·log(N) method for Ewald sums in large systems, *J. Chem. Phys.*, 1993, **98**, 10089–10092.

52 M. Cuendet and W. van Gunsteren, On the calculation of velocity-dependent properties in molecular dynamics simulations using the leapfrog integration algorithm, *J. Chem. Phys.*, 2007, **127**, 184102.

53 J. Meza, Steepest descent, *Wiley Interdiscip. Rev. Comput. Stat.*, DOI:10.1002/wics.117.

54 S. Nosé, A molecular dynamics method for simulations in the canonical ensemble, *Mol. Phys.*, 1984, **52**, 255–268.

55 W. G. Hoover, Canonical dynamics: Equilibrium phase-space distributions, *Phys. Rev. A*, 1985, **31**, 1695–1697.

56 M. Parrinello and A. Rahman, Crystal Structure and Pair Potentials: A Molecular-Dynamics Study, *Phys. Rev. Lett.*, 1980, **45**, 1196–1199.

57 D. V. D. Spoel, E. Lindahl, B. Hess, G. Groenhof, A. E. Mark and H. J. C. Berendsen, GROMACS: Fast, flexible, and free, *J. Comput. Chem.*, 2005, **26**, 1701–1718.

58 A. Phan, H. Schlösser and A. Striolo, Molecular mechanisms by which tetrahydrofuran affects CO2 hydrate Growth: Implications for carbon storage, *Chem. Eng. J.*, 2021, **418**, 129423.

59 F. Mahmoudinobar and C. L. Dias, GRADE: A code to determine clathrate hydrate structures, *Comput. Phys. Commun.*, 2019, **244**, 385–391.